\documentstyle[aps,prl,epsfig,float,twocolumn]{revtex}
\begin{document}
\twocolumn[\hsize\textwidth\columnwidth\hsize\csname
@twocolumnfalse\endcsname

\title{Noncollinear Ferromagnetism in (III,Mn)V Semiconductors}

\author{John Schliemann and A. H. MacDonald}

\address{Department of Physics, The University of Texas, Austin, TX 78712}

\address{Department of Physics, Indiana University, Bloomington, IN 47405}

\date{\today}

\maketitle

\begin{abstract}

We investigate the stability of the collinear ferromagnetic state 
in kinetic exchange models for (III,Mn)V semiconductors with randomly 
distributed Mn ions . Our results 
suggest that {\em noncollinear ferromagnetism} is commom to these
semiconductor systems. The instability of the collinear state
is due to long-ranged fluctuations involving a large fraction of
the localized magnetic moments.
We address conditions that favor the occurrence 
of noncollinear groundstates and discuss unusual behavior that we 
predict for the temperature and field dependence of its saturation 
magnetization.

\end{abstract}
\vskip2pc]

In systems with local moments, interactions that favor parallel spin 
alignment ({\em ferromagnetic} interactions) yield ferromagnetic ground states.
Interactions that favor antiparallel spin alignment ({\em antiferromagnetic} 
interactions), on the other hand, yield antiferromagnetism only in the special
case of short-range interactions on bipartite lattices. More generally,  
longer range antiferromagnetic interactions often lead to
complex order or, when spatial disorder is also present, to 
spin-glass ground states \cite{Yosida}.

In this context, the occurrence of 
ferromagnetism\cite{expgen} in III-V semiconductors with 
Mn substituted on a relatively small fraction of the cation sites is unusual, 
since Mn provides $S=5/2$ local moments in this system and the distribution 
of Mn ions among the cation sites is thought to be random.  
A ground state with parallel alignment
can be understood in a system with randomly located local moments only if
interactions are dominantly ferromagnetic, a property not normally
associated with the carrier-mediated interactions thought\cite{meanfield} to 
be responsible for ferromagnetism in
these systems.  Indeed there is ample evidence that randomness can play an 
important role in these ferromagnets, since transport and magnetic 
properties depend strongly on the MBE growth conditions used for materials 
synthesis, with reproducible results achieved only if growth protocols are
followed carefully \cite{expgen}. 

An important step toward the systematic
analysis of such effects was taken very recently by Potashnik {\it et al.}
\cite{Schiffer} who studied the effect of annealing procedures on magnetic
and transport properties for samples with a Mn fraction close to that at 
which the maximum Curie temperatures ($T_c$) occurs, $x \sim 0.05$.
They find that $T_c$, the ground 
state saturation magnetization $M(T=0)$, and the shape of the $M(T)$ curve 
all depend on the temperature and duration of annealing.  
Since the Mn content of the system is not believed to be changed by
annealing, changes in the $T=0$ magnetization can be explained only 
if Mn moments are not always, or perhaps never, completely aligned 
in the ground state.  Moreover, drastic changes in the shape of the $M(T)$ 
curve are evidence of changes in the magnetic excitation spectrum.  
Both changes most likely reflect the sensitivity of carrier-mediated 
interactions between the Mn local moments to the defect distribution that is 
varied by annealing.  

In this Letter we study the stability of states with parallel 
Mn local moments using the kinetic-exchange model of carrier-induced 
ferromagnetism.  We find that fully aligned ferromagnetic ground states do 
not occur 
when the band electrons wavefunctions are strongly concentrated around Mn ion 
locations, and propose that this circumstance must be avoided to achieve  
high ferromagnetic transition temperatures. 
On the other hand we 
find, consistent with the observations of Potashnik {\it et al.}, 
that noncollinear ferromagnetism is common in (III,Mn)V semiconductors. 
On the basis of calculations of external-field dependent magnetic excitation 
spectra, we discuss unusual temperature and field dependence
of the magnetization that we predict will be characteristic of non 
collinear states in these ferromagnets.

Ferromagnetism in (III,Mn)V semiconductors has been addressed using
first-principles local-spin-density approximation calculations
\cite{theo-lsda}, phenomenological impurity-band models\cite{bhatt} and, 
most-commonly\cite{meanfield,spinwave,mc,dassarma}, using a 
semi-phenomenological approach that sidesteps atomic scale physics by 
using an envelope function description of the bands.  The last approach is 
strongly supported by the large literature on the closely related 
(II,Mn)VI semiconductors \cite{dmsreviews} and we follow it  
here.  The model Hamiltonian we study,    
\begin{equation}
{\cal H}=\sum_{i}\frac{\vec p_{i}\,^{2}}{2m}
+\sum_{I}\int\,d^{3}r\,J(\vec r-\vec R_{I})\vec s(\vec r\, ) \cdot \vec S_{I}
\, .
\label{defmod1}
\end{equation} 
describes noninteracting carriers in a parabolic band 
characterized by an effective mass $m$, whose spin density
$\vec s(\vec r\, )$ is coupled to ionized Mn acceptor spins $\vec S_{I}$
at locations $\vec R_{I}$ by an spatially extended antiferromagnetic 
exchange coupling $J(\vec r\, )$, which we take to be of the form
\begin{equation}
J(\vec r \,)=\frac{J_{pd}}{(2\pi a_{0}^{2})^{\frac{3}{2}}}
e^{-\frac{r^{2}}{2 a_{0}^{2}}}\,,
\label{excpl}
\end{equation}
where $J_{pd}$ parametrizes the strength and $a_{0}$ the range of the 
interaction.
Similar models have been studied some time ago\cite{Abrikosov} in connection 
with classical spin glass systems; in that case however the density of 
magnetic impurities $N$ is
taken to be much smaller than the density of charge carriers $n$,
$n/N\gg 1$ \cite{Abrikosov}.  (III,Mn)V semiconductors are in the opposite, 
$n/N\ll 1$ because of strong compensation of Mn acceptors by antisite defects 
$N$.
The model (\ref{defmod1}) 
is unrealistic in its use of a single parabolic band to 
represent the semiconductor valence bands, in its neglect of Coulomb 
interactions between holes in the valence band
and other holes and Mn acceptors, and in the absence of any representation 
of the defects (other than the Mn ions) whose
distributions presumably vary during the annealing process. 
We return to these limitations of the model later.  In addition, the 
range of the exchange interaction in this
model accounts \cite{isolatedMn} for the atomic scale physics only crudely.
We believe, nevertheless, that this relatively simple model 
captures much of the physics associated with the 
approach of and the occurrence of reduced saturation magnetization states 
for non-optimal defect configurations.

Recently a theory of magnetic fluctuations around a fully aligned state
of magnetic ions was developed using the {\em virtual crystal} 
approximation \cite{spinwave}.
In contrast to that work 
we do not adopt this simplifying approximation in 
which the Mn magnetic distribution is replaced by a continuum, 
retaining individual spins of length $S=5/2$ placed at 
arbitrary locations instead.
The virtual {\em crystal approximation}, 
which has a long and successful history 
in analyzing the properties of (II,Mn)VI ferromagnets, completely 
eliminates disorder and, as we discuss further below, noncollinear 
ferromagnetism.

We first address the stability
of the collinear state in which all Mn moments have parallel orientation
taken to be in the $\hat z$ direction,
parameterizing changes in the orientation of spin $I$ by a complex number,
$z_I=(S^{x}_{I}+iS^{y}_{I})/\sqrt{2S}$. 
The zeroth-order band electron Hamiltonian is given by Eq.~(\ref{defmod1}) with
${\vec S}_I = S \hat z$.  For small changes in spin orientation 
the change in the band electron Hamiltonian is 
\begin{equation}
{\cal H}_{1}=\frac{1}{2}\sum_{I}\left[J(\vec r-\vec R_{I})
\left(\begin{array}{cc}
-z_{I}\bar z_{I} & \sqrt{2S}\bar z_{I} \\
\sqrt{2S}z_{I} & z_{I}\bar z_{I}.
\end{array}\right)
\right]
\label{perturb}
\end{equation}
The contribution to the carrier groundstate energy of first order in
the $z_{I}$ vanishes. At second order one obtains
\begin{equation}
E^{(2)}=\sum_{I,J}\bar z_{I}M_{IJ}z_{J}
\label{flucaction}
\end{equation}
where $M_{IJ}=L_{IJ}+K_{IJ}$ and  
\begin{eqnarray}
L_{IJ} & = &-\delta_{IJ}
 \int d^{3}rJ(\vec r-\vec R_{I})\langle s^{z}(\vec r)\rangle
\label{L}\\
K_{IJ} & = & \frac{S}{2}\sum_{\alpha,\beta}\Biggl[
\frac{n_{F}(\xi_{\alpha})-n_{F}(\xi_{\beta})}
{\xi_{\alpha}-\xi_{\beta}}
F^{\alpha\downarrow,\beta\uparrow}_{I}
F^{\beta\uparrow,\alpha\downarrow}_{J}\Biggr]
\label{K}
\end{eqnarray}
Here $\langle\vec s(\vec r)\rangle$ is the band-electron spin density in the 
collinear state, $n_{F}$ is the Fermi function at zero temperature
In these equations we have defined 
\begin{equation}
F^{\alpha\sigma,\beta\mu}_{I}=\int d^{3}rJ(\vec r-\vec R_{I})
\bar\psi_{\alpha\sigma}(\vec r)\psi_{\beta\mu}(\vec r)
\label{F}
\end{equation}
with $\psi_{\alpha\sigma}(\vec r)$ being the spin $\sigma$ component of the 
the carrier wave function with energy $\varepsilon_{\alpha}=
\xi_{\alpha}+\mu$, where $\mu$ is the Fermi energy. 
All eigenvalues of $M$ must be positive for the collinear state to be stable.
The diagonal contribution to $M$ from the positive definite matrix (\ref{L}) 
gives the energy change
associated with reorienting each localized spin in the mean-field created by 
the polarized 
band electrons, while the contributions from the matrix (\ref{K}) 
represents the energy reduction due to the response of band electrons to 
reorientation of the localized spins.
When the Mn spins are treated quantum mechanically using a 
Holstein-Primakoff boson representation \cite{Auerbach} and 
coherent state path integrals the $z_I$ are boson coherent state labels, 
eqivalent to replacing $z_I$ and $\bar z_{I}$ by bosonic creation and
annihilation operators $a_{I}$ and $a^{+}_{I}$. Then Eq.(~\ref{perturb}) is the 
spin-wave Hamiltonian,
and the eigenvalues\cite{retarded} of $M_{IJ}$ are the system's 
spin-wave energies. 
\begin{figure}
\centerline{\includegraphics[width=8cm]{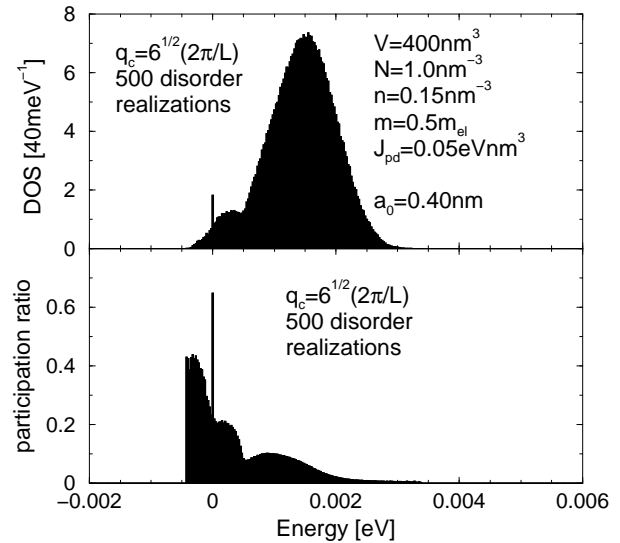}}
\caption{Upper panel:
Disorder-averaged density of states for spin-wave excitations. 
These results were obtained with a simulation cube of volume
$V=L^{3}=400{\rm nm}^{3}$ with a Mn density of $N=1.0{\rm nm}^{-3}$, 
a carrier density $n=0.15{\rm nm}^{-3}$, and $m=0.5 m_{el}$.
The strength of the exchange interaction
between ions and carriers is $J_{pd}=0.05{\rm eVnm}^{3}$ with a spatial
range of $a_{0}=0.40{\rm nm}$. 
Lower panel: Disorder-averaged participation ratios for the same system.
In both graphs the value at zero energy is enhanced 
due to the contribution of the uniform rotation mode which occurs for any
disorder realization. 
\label{fig1}}
\end{figure}

We have evaluated the matrix $M_{IJ}$ and its spectrum by solving 
for the band electron mean-field eigenstates, applying periodic boundary 
conditions to cubic simulation cells.
The single-particle wavefunctions $\psi_{\alpha\sigma}(\vec r)$ are computed
in a plane-wave basis taking into account wavevectors $\vec q$ with length up 
to an appropriate cutoff $q_{c}$.
The upper panel of Fig.~\ref{fig1} shows typical results
for the eigenvalue distribution of $L_{IJ}$ averaged over
a large number of randomly chosen Mn distributions. 
We have established by further calculations that the effects of the wavevector 
cutoff on the low-lying excitations have already
saturated for the value of $q_{c}$ used here. 
We note that for any arrangement of the Mn positions $\vec R_{I}$ the 
matrix $M_{IJ}$ contains a zero eigenvalue corresponding to a uniform 
rotation of all spins where all components $z_{I}$ are the same. This implies
summing over all columns of $M$ gives the zero vector, a sum rule that
is perfectly fulfilled in our numerics independently of 
$q_{c}$ since this plane wave cutoff does not influence the rotational 
invariance in spin space.

In the above calculations the Mn positions were chosen 
completely at random with a uniform distribution, while in a real
(III,Mn)V semiconductor the Mn ions are located on
the fcc lattice of cation sites.  Our disorder-averaged
data for the two cases is are indistinquishable, however, establishing that 
atomic-scale correlations in Mn locations do not alter our results.
Our most important finding, illustrated in Fig.~\ref{fig1} and discussed 
at length below, is that that for a wide range of parameter values 
{\em there are always a few negative eigenvalues, i.e. that the collinear 
ferromagnetic state is not stable.}

To analyze the nature of this instability we have evaluated 
participation ratios for these elementary excitations which we define by 
\begin{equation}
p_{j} =\left[NV\sum_{I}|\alpha^{j}_{I}|^{4}\right]^{-1}
\label{partrat}
\end{equation}
where $\alpha^{j}_{I}$ is the $I$-th component of the $j$-th normalized 
eigenvector of $M_{IJ}$. $p_{j}$ is an estimate for the fraction of 
Mn sites that have important involvement in
the $j$-th spin wave. For instance, if
a vector contains exactly a fraction of $p$ nonzero components of equal
modulus and all others being zero, its participation ratio is $p$.
The largest participation ratio of unity is achieved for the zero-energy
uniform rotation mode where all components of the corresponding
eigenvector are equal.

The lower panel of 
Fig.~\ref{fig1} shows the disorder-averaged participation ratio as a 
function of spin-wave energy for the same situation as in the top panel.
The property that negative energy excitations have large 
participation ratios shows that the instabilities of the collinear state 
involve correlated reorientations of many spins, rather than lone loosely 
coupled moments.  It is the high energy excitations, for which the 
mean-field term dominates, that are single-ion in character
with participation ratios comparable to $1/400$ where $400$ is the number of 
Mn ions in these simulations cells.  
The shape of the spin-wave density-of-states is, in the model we have studied, 
sensitive to the Mn density $N$, the carrier density $n$, and to the 
Hamiltonian parameters $m$, $J_{pd}$, $a_{0}$.  Situations in which the 
collinear ferromagnetic state is, for certain disorder realizations,
stable can be approached most simply (for technical reasons) 
by letting $a_{0}$ be larger, but also we believe for larger $n/N$.
For the value of $n/N$ illustrated in Fig.~\ref{fig1}, 
which corresponds to a carrier density somewhat smaller than 
that measured in the highest $T_c$ samples, 
negative eigenvalues occur for nearly any Mn ion distribution.  
We note that negative eigenvalues increase in number as the 
wavevector cutoff is increased toward its converged value.  
These results suggest that noncollinear ferromagnetic states are common, 
and that they are sensitive to the distribution of Mn ions and defects,
especially those that trap carriers.  We find that the collinear ferromagnetic 
state tends to become unstable as mean-field band eigenfunctions become 
more strongly localized around Mn ion sites.  

The effect of a weak external magnetic field on the eigenvalue spectrum
of  $M_{IJ}$ is particularly simple. The field 
couples to the local moment through its Land\'e g-factor, 
adding $2 \mu_B H$ to the energy of a spin-wave
excitation for $S=5/2$, and to the spin and orbital degrees of freedom 
of the band electrons.  The orbital coupling leads to Landau levels 
that do not play an important role at weak fields in these highly disordered 
samples. Zeeman coupling is 
also unimportant, because it gives a contribution to the 
spin-splitting that is 
negligible compared to the mean-field exchange splitting 
$\Delta=J_{pd}NS\sim 0.1 {\rm eV}$.

Thus, provided that the modulus of the smallest eigenvalue
(i.e. the onset of the eigenvalue distribution of $M_{IJ}$ at negative
energies) is small compared to $\Delta$ (as it is the case in data shown 
above), this modulus is the 
value of $g_L \mu_B H$ necessary to force full spin alignment of a 
noncollinear state. The experiments of Potashnik {\it et al.} demonstrate 
that the maximum value of $M(T=0)$ is achieved over a certain range of 
annealing histories.  
We associate this maximum value with
the fully aligned collinear Mn configuration state; indeed the maximum 
moment per Mn is consistent with full alignment partially compensated by 
band electrons.  In this circumstance, the {\em virtual crystal approximation}
used in earlier spin-wave calculations which neglected 
disorder\cite{spinwave} can be accurate.  
For other annealing histories, $M(T=0)$ is reduced
corresponding, we propose, to noncollinear order of the Mn spins.   
Calculations like those described above show that these spins gradually 
align as an external field is added to the Hamiltonian.  
Since full alignment is achieved at a finite external field, we expect full 
alignment to be indicated experimentally by a kink in the 
$M(T=0,H)$ curve.  At this point, we predict that
the system will still have gapless excitations and power-law temperature 
dependence of the magnetization, in sharp contrast to the gapful excitations 
and exponentially suppressed temperature dependence of 
conventional ferromagnets in an external
magnetic field.  At the same time, spin-resonance experiments will see a 
gapped spin-wave spectrum since they couple only to the
uniform-rotation mode which has finite energy $g_L \mu_B B$ 
over the noncollinear groundstate.
 
To address the noncollinear state, we generalize our formalism by expanding 
around and  
defining Holstein-Primakoff bosons \cite{Auerbach} 
with respect to a general  noncollinear 
orientation configuration, of each Mn spin,
${\vec S}_{I}=S\hat\Omega_{I}=S(\sin\vartheta_{I}\cos\varphi_{I},
\sin\vartheta_{I}\sin\varphi_{I},\cos\vartheta_{I})$.
If this configuration is not an extremum, there will be an energy correction 
that is {\em linear} in the Holstein-Primakoff variables:
\begin{equation}
E^{(1)}=\frac{1}{2}\sum_{I}
\left[\bar g_{I}z_{I}+g_{I}\bar z_{I}\right]
\label{gradientfluc}
\end{equation}
with $g_{I}=g^{1}_{I}+ig^{2}_{I}$, and
\begin{eqnarray}
g^{1}_{I} & =  & \sqrt{2S}\left(\vec e_{\varphi_{I}}\times\vec e_{z}\right)
\cdot\int d^{3}r\Bigl[J(\vec r-\vec R_{I})\nonumber\\
 & & \Bigl(\left(\langle\vec s(\vec r)\rangle\cdot\vec e_{\varphi_{I}}\right)
\vec e_{\varphi_{I}}+
\left(\langle\vec s(\vec r)\rangle\cdot\vec e_{z}\right)\vec e_{z}
\Bigr)\times\vec \Omega_{I}\Bigr]\\
g^{2}_{I} & =  &\sqrt{2S}\vec e_{z}\cdot
\left(\vec e_{\varphi_{I}}\times\int d^{3}rJ(\vec r-\vec R_{I})
\langle\vec s(\vec r)\rangle\right)
\end{eqnarray} 
where $\vec e_{\varphi_{I}}=(\cos\varphi_{I},\sin\varphi_{I},0)$ and
$\vec e_{z}=(0,0,1)$.
The contributions to the carrier ground state energy quadratic in the 
Holstein-Primakoff variables can also be obtained via perturbative 
calculations similar to those for the the collinear case. 
The components $g_{I}$ represent the gradient of the
energy with respect to deviations, parameterized by the $z_{I}$, from the 
reference 
orientations.  A given orientation of Mn spins is stationary with 
respect to fluctuations
if all complex coefficients $g_{I}$ vanish. This is the case if and only if
$\int d^{3}rJ(\vec r-\vec R_{I})\langle\vec s(\vec r)\rangle$ 
is parallel with the direction $\hat\Omega_{I}$ of the local ion spin.
(The collinear ferromagnetic state is always stationary but not necessarily 
stable.)

We have employed the energy gradient expression (\ref{gradientfluc}) in a 
numerical steepest descent procedure to search for true energy minima.
Briefly, our results are as follows. In cases where the energy minimum 
found by this method is close to that of the collinear state 
(with $M(T=0)$ about
90 or more percent of the maximum value), this minimum appears to be
unique for each disorder realization. 
We can therefore be confident that we have located
the absolute ground state of our model. In situations 
where the magnetisation is reduced more substantially, however, by about 20
percent or more typically, we converge to different 
energy minima from different starting points.
In these case the model has substantial 
spin-glass character with a complex energy landscape. 
For the system shown in Fig.~\ref{fig1}, for
instance, magnetisation values at local energy minima are typically 
30 to 40 percent of the collinear state value.

We thank T. Dietl, J. K\"onig, B. Lee, F. von Oppen, N. Samarth, P Schiffer,
C. Timm, and S.-R.~E. Yang for useful discussions. 
This work was supported by the 
Deutsche Forschungsgemeinschaft, the Indiana 21st Century Fund, DARPA/ONR 
Award No. N00014-00-1-0951, and the Welch foundation.


\begin{thebibliography}{50}

\bibitem{Yosida}
For a general overview see e.g. K. Yosida, {\it Theory of Magnetism},
Springer, (1996).

\bibitem{expgen}
For reviews see
H. Ohno, Science {\bf 281},  951  (1998);
H. Ohno, J. Magn. Magn. Mater. {\bf 200},  110  (1999);
H. Ohno and F. Matsukura, Solid State Commun. {\bf 117}, 179 (2001).

\bibitem{meanfield} 
T. Dietl, A. Haury, and Y.~M. d'Aubign{\'e}, 
Phys. Rev. B {\bf 55}, R3347 (1997);
T. Dietl, H. Ohno, F. Matsukura, J. Cibert, D. Ferrand, 
Science {\bf 287}, 1019 (2000); 
T. Dietl, H. Ohno, and F. Matsukura, 
Phys. Rev. B {\bf 63}, 195205 (2001);
M. Abolfath, J. Brum, T. Jungwirth, and A.~H. MacDonald, 
Phys. Rev. B {\bf 63}, 054418 (2001).

\bibitem{Schiffer}
S.~J. Potashnik, K.~C. Ku, S.~H. Chun, J.~J. Berry, N. Samarth, 
and P. Schiffer, Appl. Phys. Lett. {\bf 19}, 1495 (2001).  
The dependence of the crystal properties of these ferromagnets 
on the growth conditions has also been 
studied recently by G.~M. Schott, W. Faschinger, and L.~W. Molenkamp,
cond-mat/0105562. 

\bibitem{theo-lsda}
H. Akai, Phys. Rev. Lett. {\bf 81}, 3002 (1998);
S. Sanvito, P. Ordejon, and N.~A. Hill, Phys. Rev. B {\bf 63}, 165206 (2001);
S. Sanvito and N.~A. Hill, Appl. Phys. Lett. {\bf 78}, 3493 (2001);
M. van Schilfgaarde and O.~N. Mryasov, Phys. Rev. B {\bf 63}, 233205 (2001).

\bibitem{bhatt}
X. Wan and R.~N. Bhatt, cond-mat/0009161;
M. Berciu and R.~N. Bhatt, Phys. Rev. Lett. {\bf 87}, 107203 (2001);
M.~P. Kennett, M. Berciu, R.~N. Bhatt, cond-mat/0102315.

\bibitem{spinwave}
J. K\"onig, H.~H. Lin, and A.~H. MacDonald, 
Phys. Rev. Lett. {\bf 84}, 5628 (2000);
cond-mat/0010471, published in {\it Interacting Electrons in 
Nanostructures}, edited by R. Haug and H. Schoeller, Springer (2001);
J. K\"onig, T. Jungwirth, and A.~H. MacDonald, 
Phys. Rev. B {\bf 64}, 184423 (2001).

\bibitem{mc}
J. Schliemann, J. K\"onig, H.~H. Lin, and A.~H. MacDonald, 
Appl. Phys. Lett. {\bf 78}, 1550 (2001);
J. Schliemann, J. K\"onig, and A.~H. MacDonald, 
Phys. Rev. B {\bf 64}, 165201 (2001).

\bibitem{dassarma} A. Chattopadhyay, S. Das Sarma, and A.~J. Millis, 
cond-mat/0106455.

\bibitem{dmsreviews} J.~K. Furdyna, J. Kossut, {\it Diluted Magnetic 
Semiconductors}, in {\it Semiconductors and Semimetals}, volume 25, 
Academic Press (1988); 
T. Dietl, {\it Diluted Magnetic Semiconductors}, in {\it Handbook of
Semiconductors}, volume 3B, North-Holland (1994).

\bibitem{Abrikosov}
For a review see A.~A. Abrikosov, Adv. Phys. {\bf 29}, 869, (1980).

\bibitem{isolatedMn} A.~K. Bhattacharjee and C. Benoit $\grave { \rm a}$ 
la Guillaume, Solid State Commun. {\bf 113}, 17 (2000);
V.~I. Litvinov and V.~K. Dugaev, Phys. Rev. Lett. 
{\bf 86}, 5593 (2001).

\bibitem{Auerbach}
See e.g. A. Auerbach, {\it Interacting Electrons and Quantum Magnetism}
Springer, (1994).

\bibitem{retarded} Strictly speaking  $K_{IJ}$ is frequency dependent in a 
quantum formulation, reflecting the retarded nature of the carrier-mediated
interaction between
local moments.  The frequency dependence is weak however, because of the 
small ratio between spin-wave excitation energies and the mean-field 
spin-splitting of the itinerant electron bands.
It can be shown that, for high participation ratio 
low frequency excitations, the only role 
of the frequency dependence is to incorporate the band electron 
contribution to the magnetization and to rescale the collective mode 
energies by a factor of $1/(1-n/2SN)$.  
Terms at second and higher order in the low frequency expansion of $K_{IJ}$
reflect correlated fluctuations between local moment and itinerant electron
spins that supress the total magnetization by a further small factor.

\end{thebibliography}
\end{document}